\documentclass[12pt]{article}

\usepackage{fullpage}
\usepackage{subfigure}
\usepackage{latexsym}
\usepackage{graphicx}
\usepackage[intlimits]{amsmath}
\usepackage{amsthm}
\usepackage{natbib}
\usepackage{amsfonts}
\usepackage{url}
\usepackage{ulem}

\DeclareMathOperator*{\argmax}{arg\,max}

\newcommand{\cprob}[2]{\ensuremath{\text{Pr}\left(#1 \,|\,#2\right)}}  
\newcommand{\prob}[1]{\ensuremath{\text{Pr}\left(#1 \right)}}
\newcommand{\cexpect}[4]{\ensuremath{\text{E}\left#3 #1 \,|\,#2\right#4}}  
\newcommand{\expect}[3]{\ensuremath{\text{E}\left#2 #1 \right#3}}

\newcommand{\compdata}{\ensuremath{\mathbf{x}}}
\newcommand{\obsdata}{\ensuremath{\mathbf{y}}}
\newcommand{\hiddata}{\ensuremath{\mathbf{h}}}
\newcommand{\compdatascalar}{\ensuremath{x}}
\newcommand{\hiddatascalar}{\ensuremath{h}}
\newcommand{\obsdatascalar}{\ensuremath{y}}
\newcommand{\truesampl}{\ensuremath{p_T}}
\newcommand{\falsesampl}{\ensuremath{p_F}}
\newcommand{\falsesamplcomp}[1]{\ensuremath{p_{F#1}}}
\newcommand{\truepar}{\ensuremath{\boldsymbol{\theta}_T}}

\newcommand{\falsepar}{\ensuremath{\boldsymbol{\theta}_F}}
\newcommand{\subpar}{\ensuremath{\boldsymbol{\theta}_0}}
\newcommand{\falseparscalar}[1]{\ensuremath{\theta_{F#1}}}
\newcommand{\falseparest}{\ensuremath{\hat{\boldsymbol{\theta}}_F}}
\newcommand{\falseparcomp}[1]{\ensuremath{\theta_{F#1}}}
\newcommand{\falseparestcomp}[1]{\ensuremath{\hat{\theta}_{F#1}}}
\newcommand{\trueparspace}{\ensuremath{\boldsymbol{\Theta}_T}}
\newcommand{\falseparspace}{\ensuremath{\boldsymbol{\Theta}_F}}
\newcommand{\subparspace}{\ensuremath{\boldsymbol{\Theta}_0}}
\newcommand{\compsum}{\ensuremath{\mathbf{s}}}
\newcommand{\compsumscalar}{\ensuremath{s}}
\newcommand{\summean}{\ensuremath{\boldsymbol{\mu}}}
\newcommand{\summeanscalar}{\ensuremath{\mu}}
\newcommand{\markovchain}[1]{\ensuremath{X_{#1}}}
\newcommand{\statespace}{\ensuremath{\mathcal{S}}}
\newcommand{\infgencomp}{\ensuremath{\lambda}}
\newcommand{\infgen}{\ensuremath{\boldsymbol{\Lambda}}}
\newcommand{\statdistcomp}{\ensuremath{\pi}}
\newcommand{\statdist}{\ensuremath{\boldsymbol{\statdistcomp}}}
\newcommand{\transprob}{\ensuremath{\mathbf{P}}}
\newcommand{\transprobscalar}{\ensuremath{p}}
\newcommand{\labelset}{\ensuremath{\mathcal{L}}}

\newcommand{\suffstat}{\ensuremath{\mathbf{t}}}
\newcommand{\suffstatscalar}{\ensuremath{t}}
\newcommand{\mixpropscalar}{\ensuremath{\alpha}}
\newcommand{\mixprop}{\ensuremath{\boldsymbol{\alpha}}}
\newcommand{\indfun}[1]{\ensuremath{1_{\{#1\}}}}

\newcommand{\asarrow}{\ensuremath{\overset{\text{a.s.}}{\rightarrow}}}

\newcommand{\conv}{plug-in\ }
\newcommand{\robust}{imputation\ }
\newcommand{\Conv}{Plug-in\ }
\newcommand{\Robust}{Imputation\ }

\newtheorem{theorem}{Theorem}
\newtheorem{corollary}{Corollary}

\begin{document}

\begin{center}
  {\LARGE Imputation Estimators Partially Correct for Model Misspecification}\\\ \\
  {Vladimir N. Minin$^{1}$, John D. O'Brien$^2$, Arseni Seregin$^1$ \\ 
    $^1$Department of Statistics, University of Washington Seattle, WA, 98195, USA \\
    $^2$Department of Mathematics, University of Bristol, Bristol, Avon, BS1, UK \\ 
  }
\end{center}

\begin{abstract}
Inference problems with incomplete observations often aim at estimating population properties of unobserved 
quantities. One simple way to accomplish this estimation is to impute the unobserved quantities of interest
at the individual level and then take an empirical average of the imputed values. We show that this simple 
imputation estimator can provide partial protection against model misspecification.
We illustrate 
imputation estimators' robustness to model specification on three examples:
mixture model-based clustering, estimation of genotype frequencies in population genetics, 
and estimation of Markovian evolutionary distances.
In the final example, using a representative model misspecification, 
we demonstrate that in non-degenerate 
cases, the imputation estimator dominates the plug-in estimate asymptotically. 
We conclude by outlining a Bayesian implementation of the imputation-based estimation.

\end{abstract}


\clearpage

\section{Introduction}
We are interested in robustness to model misspecification in problems with incomplete observations. 
Semiparametric approaches have enjoyed a lot of success in
this area but these methods lack universality and so need to be fine-tuned for each problem at 
hand \citep{TsiatisBook,Little2004,Kang2007,Chen2009}. Consequently, when practitioners are faced with nonstandard 
problems with incomplete observations, they are often left to their own devices. 
As a first step to ameliorating this deficiency, we propose a general imputation-based estimation method that 
provides partial protection against model misspecification for incomplete data problems. 
\par
The idea of using imputation techniques to combat model misspecification 
is not new. Consider the standard missing data problem of estimating population mean $\mu$
given a sample $(z_1,y_1z_1,\mathbf{x}_1),\dots,(z_n,y_nz_n,\mathbf{x}_n)$, where 
$y_i$ is a response variable, $z_i$ is a response indicator taking value 1 if $y_i$ is observed 
and 0 otherwise, and $\mathbf{x}_i$ is a vector of covariates. Assuming strong ignorability, meaning 
that $y_i$ and $z_i$ are independent given $\mathbf{x}_i$, we 
use only those individuals for which the response 
variable is available to fit a response model with $m_i = \text{E}(y_i\,|\,x_i)$ to obtain $\hat{m}_i$ 
\citep{Rosenbaum1983}.
Intuitively, we can combine the empirical estimate of the mean of respondents with model-based predictions of 
missing $y_i$s for non-respondents to arrive at
$\hat{\mu} 
   = (1/n)\sum_{i: z_i=1}^n y_i + (1/n)\sum_{i:z_i=0}^n \hat{m}_i$.
This estimator, called an imputation estimator by \citet{Tsiatis2007}, will be biased if the response model 
is misspecified.  However, the bias vanishes as the number of non-respondents decreases to zero. Using conditioning
on the observed data, we can rewrite \citet{Tsiatis2007}'s imputation estimator as 
$\hat{\mu} = (1/n)\sum_{i=1}^n \text{E}(y_i \mid z_i, y_iz_i,\mathbf{x}_i)$. In a completely unrelated missing data
setting, \citet{OBrien2009} also use 
expectations of complete data conditional on the observed data to arrive at novel estimators of evolutionary
distances. Although \citet{OBrien2009} used imputation by conditional expectations explicitly, these authors 
did not recognize the full generality of their approach.
\par
In this paper, we investigate the behavior of imputation estimators when they are 
applied to general problems with incomplete observations.
After formulating the generalized imputation estimator, we consider three problems with incomplete observations.
We start with a mixture model and demonstrate that imputation is useful for estimating densities of mixture components. 
Moreover, this imputation density estimation improves accuracy of mixture model-based clustering. Next, we turn to 
a statistical genetics problem of estimating genotype frequencies. To keep the genetic-specific intricacies to a minimum, 
we construct an artificial but representative example. In spite of the introduced simplification, 
our results are directly applicable to a topical problem of 
multilocus haplotype/genotype frequency estimation, where model misspecification occurs due to a failure
to account for population structure \citep{Allen2008,Kraft2005}. In our last example, we consider 
\robust estimators of evolutionary distances between
DNA sequences with partially observed continuous-time Markov chains introduced in \citet{OBrien2009}. 
We fill some theoretical gaps in their work. First, we
identify situations where \robust estimators are not helpful. In doing so, we - for the first time to our 
knowledge -
use the fact that so called group-based Markov models belong to the regular exponential
family \citep{Evans1993}. Next, we compute almost sure limits of \robust and \conv estimators for a particular 
model misspecification. 
Although we make several simplifying assumptions in this derivation, we believe that qualitatively our results are
portable to more realistic applications considered by \citet{OBrien2009}. We conclude by outlining a Bayesian 
implementation of the imputation-based estimation.

\section{Generalized imputation estimators}

Assume that complete data 
$\compdata = (\compdata_1,\dots,\compdata_n)$ are independent and identically distributed with each $\compdata_i$ 
distributed according to a parametric family of sampling densities 
$\truesampl(\compdata;\truepar)$ with parameters $\truepar \in \trueparspace$. We observe
each $\compdata_i$ through a transformed vector $\obsdata_i = \obsdata(\compdata_i)$.
We further assume that the true sampling density $\truesampl(\compdata_1;\truepar)$ is unknown to us and 
we have to erroneously postulate a misspecified model $\falsesampl(\compdata_1;\falsepar)$, where 
$\falsepar \in \falseparspace$ with parameter spaces $\trueparspace$ and $\falseparspace$ of possibly 
different dimensions.
Despite this model misspecification, we would like to estimate 
$\summean = \text{E}_{\truepar}[\compsum(\compdata_1)] = \int \compsum(\compdata_1) 
  \truesampl(\compdata_1;\truepar)\text{d}\compdata_1$,
where $\compsum$ is an arbitrary measurable function that maps complete data to an $m$-dimensional 
vector of summary statistics. 
Assuming that $\falsepar$ is identifiable from incomplete data $\obsdata = (\obsdata_1,\dots,\obsdata_n)$, 
one can simply maximize the likelihood of the observed data
$\prod_{i=1}^n \falsesampl(\obsdata_i;\falsepar)$ 
to arrive at the maximum likelihood estimate 
$\falseparest = \argmax_{\falsepar \in \falseparspace} \falsesampl(\obsdata;\falsepar)$.
Then, \textit{ignoring model misspecification}, we use $\falseparest$
to get the \conv estimate of the complete-data summaries
\begin{equation}
  \hat{\summean}_n^{pi} = \text{E}_{\falseparest}[\compsum(\compdata_1)] = 
  \int \compsum(\compdata_1) \falsesampl(\compdata_1;\falseparest)\text{d}\compdata_1.
  \label{conv_est_gen}
\end{equation}
This estimator is destined to be biased and asymptotically inconsistent in nearly all situations
due to the model misspecification.
Consider an \robust estimator
\begin{equation}
  \hat{\summean}_n^{im} = \frac{1}{n} \sum_{i=1}^n \text{E}_{\falseparest}
  \left[\compsum(\compdata_i) \,|\,\obsdata_i\right] = 
  \frac{1}{n} \sum_{i=1}^n 
  \int \compsum(\compdata_i) \falsesampl(\compdata_i\,|\,\obsdata_i;\falseparest)\text{d}\compdata_i.
  \label{robust_est_gen}
\end{equation}
The motivation behind this new estimator is quite simple: in order to offer protection against model 
misspecification, we would like to use the empirical measure based on $\obsdata_1,\dots,\obsdata_n$. 
To accomplish this, we write
$  \text{E}_{\truepar}\left[\compsum(\compdata_1)\right] = 
  \text{E}_{\truepar}\left\{\text{E}_{\truepar}\left[\compsum(\compdata_1)\,|\,\obsdata_1\right]\right\}
  \approx
  \mathbb{P}_n\text{E}_{\truepar}\left[\compsum(\compdata_1)\,|\,\obsdata_1\right]$
where $\mathbb{P}_n f = \frac{1}{n}\sum_{i=1}^n f(\obsdata_i)$ for any measurable function $f$.
In the absence of a good alternative, we plug-in $\falseparest$ for $\truepar$ in the conditional expectations 
of $\compsum(\compdata_i)$ to arrive at our \robust estimator, $\hat{\summean}_n^{im}$.
\par
If the family of distributions $\{\falsesampl(\obsdata;\falsepar)\}$ satisfies usual regularity conditions we have
$\falseparest\asarrow\subpar$. 
For example, if our model is not misspecified, i.e. $\falseparspace\equiv\trueparspace$, we would have 
$\subpar = \truepar$. 
Consider the family of functions $\mathcal{F}$ which consists of conditional expectations:
$\mathcal{F} = \{f(\obsdata_1;\boldsymbol{\theta})=\text{E}_{\boldsymbol{\theta}}
  \left[\compsum(\compdata_1)\,|\,\obsdata_1)\right], \boldsymbol{\theta}\in\subparspace\}$
for some bounded open neighborhood $\subparspace$ of the limiting value $\subpar$. If we assume that $\mathcal{F}$
has finite bracketing number $N_{[]}(\varepsilon,\mathcal{F},L_{1}(P))$ for each $\varepsilon>0$ and 
is pointwise continuous in $\boldsymbol{\theta}$, then one can show that
$\mathbb{P}_n \text{E}_{\falseparest}\compsum\left[\compdata_1)\,|\,\obsdata_1\right] \asarrow 
\text{E}_{\truepar}\left\{\text{E}_{\subpar}\left[\compsum(\compdata_1)\,|\,\obsdata_1\right]\right\}$
using standard empirical processes machinery \citep{vanderVaartWellnerBook}. Assuming model misspecification 
almost inevitably leads to $\subpar \ne \truepar$. Therefore, our \robust estimator has little chance of 
achieving asymptotic consistency. However, if the fraction of missing information is relatively small, 
our new estimator can be quite close to the true value both for finite sample sizes and asymptotically.
\par
Assume that a misspecified complete-data sampling density belongs to the regular exponential 
family so that $\falsesampl(\compdata_1;\falsepar) = a(\compdata_1) 
  \exp\left[\falsepar^T \suffstat(\compdata_1) \right]/b(\falsepar)$,
where $\suffstat(\compdata_1)=(\suffstatscalar_1(\compdata_1),\dots,\suffstatscalar_r(\compdata_1))$ 
is an $r$-dimensional vector of minimal sufficient statistics and 
$\falsepar = (\falseparscalar{1},\dots,\falseparscalar{r})$ is a natural parameter vector of the 
same dimension.
Then, as noted by \citet{Sundberg1974}, the likelihood equations based on the observed data 
$\obsdata$ can be written as 
$(1/n)\sum_{i=1}^n \text{E}_{\falsepar}\left[\suffstat(\compdata_i)\,|\,\obsdata_i\right] 
  = \text{E}_{\falsepar}\left[\suffstat(\compdata_1)\right]$.
Therefore, if the complete-data summary 
$\compsum(\compdata_1)$ can be expressed as a linear transformation of the sufficient statistics 
$\suffstat(\compdata_1)$, imposed by the falsely assumed regular exponential family model, 
then the \conv estimator (\ref{conv_est_gen}) and \robust estimator 
(\ref{robust_est_gen}) \textit{coincide exactly} regardless of the true sampling density of
$\compdata_1$.

\section{Mixture models and model-based clustering}
Consider a mixture model with $k$ components. Let $\hiddata = (\hiddatascalar_1,\dots,\hiddatascalar_n)$ 
be iid discrete random variables taking values 
in $\{1,\dots,k\}$ with probabilities $\prob{\hiddatascalar_1 = j} = \mixpropscalar_j$, $\sum_{j=1}^k 
\mixpropscalar_j=1$. Event $\hiddatascalar_i = j$ indicates that the observed $\obsdata_i$ is sampled from the density 
$\falsesamplcomp{j}(\obsdata;\falseparcomp{j})$. 
The complete-data sampling density becomes
\begin{equation*}
  \falsesampl(\hiddatascalar_i,\obsdata_i;\falsepar) = \prod_{j=1}^k 
  \left[\mixpropscalar_j \falsesamplcomp{j}(\obsdata_i;\falseparcomp{j})\right]^{\indfun{\hiddatascalar_i=j}}
\end{equation*}
We obtain parameter estimates $\hat{\mixprop} = (\hat{\mixpropscalar}_1,\dots,\hat{\mixpropscalar}_k)$ 
and $\falseparest = (\falseparestcomp{1},\dots,\falseparestcomp{k})$
by maximizing $\prod_{i=1}^n\falsesampl(\obsdata_i;\falsepar)$, where
$\falsesampl(\obsdata_i;\falsepar) = \sum_{j=1}^k \mixprop_j \falsesamplcomp{j}(\obsdata_i;\falseparcomp{j})$.
If we further assume regular exponential family sampling densities of mixture components sharing the same 
normalizing constant $a(\obsdata)$,
$\falsesamplcomp{j}(\obsdata;\falsesampl) =
  a(\obsdata)\exp\left[\suffstat_j(\obsdata)^T\falseparcomp{j} \right]/b_j(\falseparcomp{j})$,
then the density of the $i$th completely observed sampling unit also belongs to the regular exponential family,
\begin{equation*}
  \falsesampl(\hiddatascalar_i,\obsdata_i;\falsepar) = a(\obsdata_i)
  \exp\left\{ \sum_{j=1}^k \indfun{\hiddatascalar_i=j}\suffstat_j(\obsdata_i)^T\falseparcomp{j} + \sum_{j=1}^k 
    \indfun{\hiddatascalar_{i}=j} \left[\ln \frac{\mixpropscalar_j}{b_j(\falseparcomp{j})}\right] \right\}.
\end{equation*}
From our discussion of regular exponential family complete-data likelihoods, 
it is clear that \conv and \robust estimators of mean complete-data summaries,
\begin{equation}
  \text{E}_{\truepar}\left[\indfun{\hiddatascalar_i=j} \suffstat_j(\obsdata_1)\right] \text{ and }
  \text{E}_{\truepar}\left[\indfun{\hiddatascalar_i=j}\right],
\label{mix_suff_stat}
\end{equation}
will coincide exactly regardless of the true complete-data sampling model 
$\truesampl(\obsdata_1,\hiddata_1;\truepar)$.
In fact, \conv and \robust estimators of the second mean complete-data summary, 
$\text{E}_{\truepar}\left[\indfun{\hiddatascalar_i=j}\right]$, will coincide even if densities
$\falsesamplcomp{j}(\obsdata_i;\falsepar)$ do not belong to the regular exponential family. 
To see this, note that the \conv estimator in this context is
$\hat{\mixpropscalar}^{pi}_j = \text{E}_{\hat{\mixpropscalar}_j}\left[\indfun{\hiddatascalar_i=j}\right] = 
  \text{Pr}(\hiddatascalar_i=j) = \hat{\mixpropscalar}_j$.
The estimated probability that observation $i$ belongs to component $j$ is
\[
\hat{z}_{ij} = \cexpect{\indfun{\hiddatascalar_i=j}}{\obsdata_i}{(}{)} =  
\frac{\hat{\mixpropscalar} \falsesamplcomp{j}(\obsdata_i,\falseparestcomp{j})}
{\sum_{j=1}^k \hat{\mixpropscalar} \falsesamplcomp{j}(\obsdata_i,\falseparestcomp{j})}.
\]
The \robust estimate of the $j$th mixing proportion becomes
$\hat{\mixpropscalar}_j^{im} = (1/n) \sum_{i=1}^n \cexpect{\indfun{\hiddatascalar_i=j}}{\obsdata_i}{(}{)} = 
  (1/n) \sum_{i=1}^n \hat{z}_{ij}$.
The likelihood equations for the mixture model can be rearranged to 
show that $\hat{\mixpropscalar}^{pi}_j=\hat{\mixpropscalar}^{im}_j$ \citep{Redner1984}.
Notice that estimating all of the above complete-data expectations requires unambiguously 
identifying mixture component $j$, which we assume is possible by imposing constraints on
mixture component parameters $\falseparcomp{1},\dots,\falseparcomp{k}$.
\par
\begin{figure}
  \includegraphics[height=\textwidth, angle=-90]{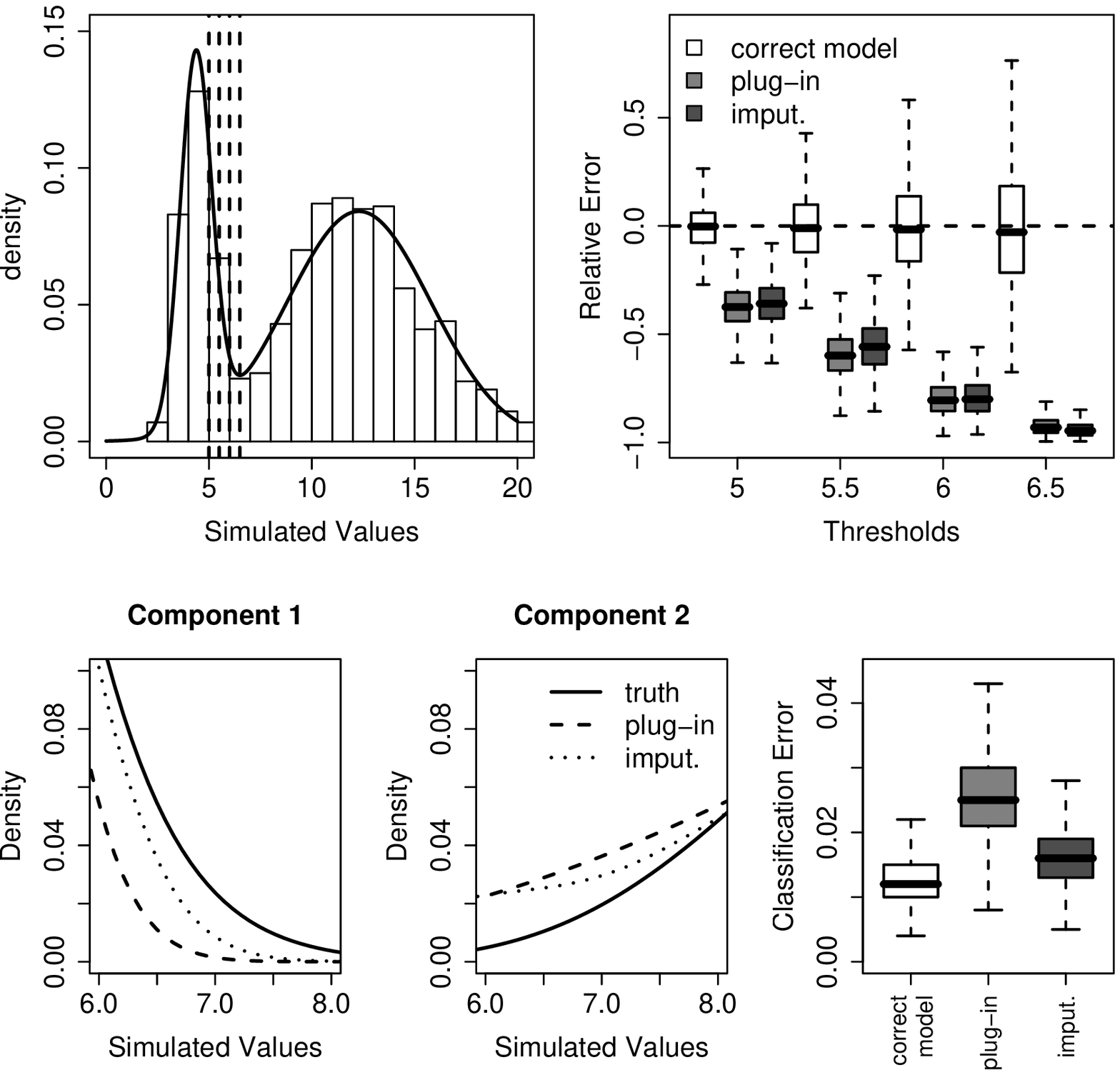}
  \caption{\small Mixture model example. The upper left plot shows a histogram of 1000 simulated realizations of the 
    two-component log-normal model, described in the text. The solid line depicts the normal mixture density
    estimated from these simulated data. The dashed vertical lines indicate four values of threshold $c$, for 
    which we estimate 
    $\summeanscalar(c) = \text{E}_{\truepar}\left(\indfun{\hiddatascalar_1=1} \indfun{\obsdatascalar > c}\right)$.
    Results of conventional and robust estimation of these quantities are shown in the upper right plot of the figure.
    We repeat simulation and estimation 1000 times and plot box-plots of relative errors, 
    $\frac{\hat{\summeanscalar}^{pi}(c)-\summeanscalar(c)}{\summeanscalar(c)}$ and 
    $\frac{\hat{\summeanscalar}^{im}(c)-\summeanscalar(c)}{\summeanscalar(c)}$, for $c=5.0, 5.5, 6.0, 6.5$.
    The bottom row shows results of mixture component density estimation and classification errors during 
    model based clustering.
  }
  \label{normal_mix}
\end{figure}

To make our discussion of mixture models more concrete, 
we simulate $n=1000$ realizations from a mixture of two 
log-normal distributions with the log-scale means $\mu_1=1.5$ and $\mu_2=2.5$ and standard deviations
$\sigma_1 =0.2$ and 
$\sigma_2=0.25$ respectively. The mixing proportion, $\alpha$, was set to $0.3$, completing
the set of true model parameters $\truepar = (\mu_1,\mu_2,\sigma_1,\sigma_2,\alpha)$.
Now, we assume a two-component normal mixture model with means $\nu_1$, $\nu_2$, possibly unequal standard deviations 
$\delta_1$, $\delta_2$, and a mixing proportion $\beta$. We estimate parameters 
$\falsepar = (\nu_1,\nu_2,\delta_1,\delta_2,\beta)$ of this misspecified model using maximum 
likelihood via the EM algorithm \citep{Dempster1977,Fraley2003}. We show a histogram of simulated data with 
a normal mixture model fit in the left plot of Figure~\ref{normal_mix}.
\par
To avoid the label switching problem, we define 
mixture component labels by the inequality $\nu_1 < \nu_2$. 
Equation (\ref{mix_suff_stat}) says that
if we try to estimate $\text{E}_{\truepar}\left(\indfun{\hiddatascalar_1=1} \obsdatascalar_1\right)$, 
$\text{E}_{\truepar}\left(\indfun{\hiddatascalar_1=1} \obsdatascalar_1^2\right)$ or 
$\text{E}_{\truepar}\left(\indfun{\hiddatascalar_1=1}\right)$, 
it does not matter whether we use the \conv or 
\robust approach. Instead, we choose to estimate the proportion of samples from the first mixture component
that fall to the right of some threshold $c$, 
$\summeanscalar(c) = \text{E}_{\truepar}\left(\indfun{\hiddatascalar_1=1} \indfun{\obsdatascalar_1>c}\right)$.
The \conv estimate of this quantity is
\begin{equation}
  \hat{\summeanscalar}^{pi}(c)= \text{E}_{\falseparest}\left(\indfun{\hiddatascalar_1=1} \indfun{\obsdatascalar_1>c}\right)= 
  \left[1-\Phi\left(\frac{c-\hat{\nu}_1}{\hat{\delta}_1}\right)\right]\hat{\beta},
  \label{conv_est_example}
\end{equation}
where $\Phi$ is the standard normal cdf. Our \robust estimator becomes
\begin{equation*}
  \hat{\summeanscalar}^{im}(c) = \frac{1}{n}\sum_{i=1}^n 
  \text{E}_{\falseparest}\left(\indfun{\hiddatascalar_i=1} 
    \indfun{\obsdatascalar_i>c}\,|\,\obsdatascalar_i\right) = 
  \frac{1}{n}\sum_{i=1}^n \hat{z}_{ij} \indfun{y_i > z}.
\end{equation*}
Since tails of mixture components can be estimated via imputation, it should be possible to devise an 
\robust estimator of mixture components' densities. Indeed, if we use a nonparametric kernel density estimator, where
each observed point $i$ is weighted by $z_{ij}$, we arrive at an \robust estimate of the $j$th component density.
This is potentially useful, because more accurate estimation of component densities may lead to more accurate
model-based clustering \citep{Fraley2002}.
\par
The right plot of Figure~\ref{normal_mix} demonstrates results of estimating $\summeanscalar(c)$ for 
threshold values $c = 5.0, 5.5, 6.0, 6.5$, depicted in the left plot of the figure by the dashed 
vertical lines. 
We consider these values of $c$, because they fall into the region where sampled points 
can not be easily assigned to either of the two mixture components. We simulate $1000$ test data sets
using already described settings. For each of the simulated data set, we compute \conv 
estimates of $\summeanscalar(c)$ using the fitted correct log-normal and the misspecified normal model and
the \robust under the misspecified normal model. We show box plots of the corresponding relative 
errors in the upper right plot of Figure~\ref{normal_mix}. Although, the 
performance of the \conv and \robust estimators under model misspecification is disappointingly similar, 
\robust density estimates, plotted in the bottom row, look more promising. We used \conv density estimates
under the correct and misspecified model and \robust density estimates to assign simulated points to two
clusters. We then computed clustering classification error using R package MCLUST \citep{Fraley2003}.
As shown in the lower bottom plot, clustering accuracy improves significantly 
under \robust estimates of mixture component densities and approaches the accuracy of clustering under the 
correct mixture model.
\par

\section{Estimating genotype frequencies}

Here, we turn to a classical problem in statistical genetics: estimating allele and 
genotype frequencies from incomplete
observations \citep{Ceppelini1955}. Suppose that we measure some observable characteristic, called a phenotype,
in $n$ individuals and record them in a vector $\obsdata = (\obsdatascalar_1,\dots,\obsdatascalar_n)$, where
each $y_i$ takes one of $M$ possible values in $\mathcal{C} = \{c_1,\dots,c_M\}$. 
We further assume that each individual
$i$ has an unobserved genotype $\compdata_i = (\compdatascalar_{i1},\compdatascalar_{i2})$, 
defined as an unordered pair of gene variants, called alleles, on two paired chromosomes of this individual. 
Suppose there are $R$ possible alleles, $\mathcal{G}=(g_1,\dots,g_R)$. Genotypes are assumed to 
determine observed phenotypes via a deterministic function 
$h: \mathcal{G} \times \mathcal{G} \rightarrow \mathcal{C}$ such that $h(g_k,g_l)  = h(g_l,g_k)$. 
Making certain population genetics assumptions allows us to assume that unobserved genotypes are iid with 
\begin{equation}
\truesampl((g_k,g_l);\mathbf{p},f) = 
\begin{cases}
  p_k^2(1-f) + fp_k &\text{ if } k = l\\
  2p_k p_l(1-f) &\text{ if } k \ne l,
\end{cases}
\label{inbreed-model}
\end{equation}
where $\mathbf{p} = (p_1,\dots,p_R)$ are population allele frequencies and $f$ is called an inbreeding coefficient. 
We erroneously assume that $f=0$, reducing the model for genotype probabilities to 
the celebrated Hardy-Weinberg equilibrium \citep{Hardy1908,Weinberg1908}. 
The falsely misspecified complete-data likelihood for datum 1 becomes
\[
\falsesampl(\compdata_1;\mathbf{p}) = \prod_{k > l} \left(2 p_k p_l\right)^{\indfun{\compdata_1=(g_k,g_l)}}
\prod_{k=1}^R (p_i)^{2 \times \indfun{\compdata_1=(g_k,g_k)}} 
\propto \prod_{k=1}^R p_k^{t_k},
\]
where $t_k = 2 \times \indfun{\compdata_1=(g_k,g_k)}+\sum_{l=1}^R \indfun{\compdata_1=(g_k,g_l)}$.
The misspecified observed-data likelihood for datum 1 is
$\falsesampl(\obsdatascalar_1;\mathbf{p}) = \sum_{\compdata_1:h(\compdata_1)=\obsdatascalar_1}
\falsesampl(\compdata_1;\mathbf{p})$.
\par
Since the complete-data likelihood is in the regular exponential family with sufficient statistics
$(t_1,\dots,t_R)$, the \conv and \robust estimates of $\text{E}(\sum_{i=1}^R c_it_i)$ will coincide exactly.
Suppose our objective is to estimate genotype frequencies
$\mu_{kl} = \expect{\indfun{\compdata_1 = (g_k,g_l)}}{(}{)} = 
\prob{\compdata_1 = (g_k,g_l)}$.
The complete-data summary $\indfun{\compdata_1=(g_k,g_l)}$ can not be expressed as a linear combination of 
the sufficient statistics, so \conv and \robust estimation will not necessarily produce identical results.
After obtaining maximum likelihood estimates of allele frequencies, $\hat{\mathbf{p}}$,
the \conv approach yields
\[
\hat{\mu}_{kl}^{pi} = \hat{p}_i^2 \indfun{k = l} + 2\hat{p}_k \hat{p}_l \indfun{k \neq l}.
\]
The \robust estimator becomes
\[
\hat{\mu}_{kl}^{im} = 
\frac{1}{n} \sum_{i=1}^n \cprob{\compdata_i=(g_k,g_l)}{\obsdatascalar_i=c_j} \indfun{y_i=c_j} =  
  \frac{n_j \hat{p}_{k}^2}{n \falsesampl(c_j;\hat{\mathbf{p}})} \indfun{k =l}+
  \frac{n_j 2 \hat{p}_{k} \hat{p}_l}{n \falsesampl(c_j;\hat{\mathbf{p}})} 
  \indfun{k \neq l},
\]
where $h(g_k,g_l) = c_j$ and $n_j = \sum_{i=1}^n \indfun{y_i=c_j}$.

\par
\par
\begin{table}
  \centering
  \caption{Mappings of complete to observed data during genotype frequencies estimation.
  Ambiguous phenotypes are highlighted in bold.}
  \label{hw-maps}
  \medskip
  \begin{tabular}{|r|cccccccccc|}\hline
    $(g_k,g_l)$ & $(A,A)$ & $(A,B)$ & $(A,C)$ & $(A,D)$ & $(B,B)$ & $(B,C)$ &     
    $(B,D)$ & $(C,C)$ & $(C,D)$ & $(D,D)$\\
    \hline
    $h_1(g_k,g_l)$ & $aa$ & $ab$ & $ac$ & $ad$ & $bb$ & $\boldsymbol{bdc}$ & 
    $\boldsymbol{bdc}$ & $cc$ & $cd$ & $dd$ \\ 
    $h_2(g_k,g_l)$ &  $aa$ &  $ab$ & $ac$ &  $ad$ & $\boldsymbol{bd}$ & $\boldsymbol{bdc}$ &      
    $\boldsymbol{bdc}$ & $cc$ & $cd$ & $\boldsymbol{bd}$\\
    \hline
\end{tabular}
\end{table}
Consider a particular case of the above model with four alleles: $\mathcal{G} = \{A, B,C,D\}$. Table~\ref{hw-maps}
defines two mappings from genotypes to phenotypes, $h_1: \mathcal{G} \times \mathcal{G} \rightarrow \mathcal{C}_1$
and $h_2: \mathcal{G} \times \mathcal{G} \rightarrow \mathcal{C}_2$, where 
$\mathcal{C}_1 = \{aa,ab,ac,ad,bb,bdc,cc,cd,dd\}$ and $\mathcal{C}_2 = \{aa,ab,ac,ad,bd,bdc,cc,cd\}$. 
Notice that $\mathcal{C}_1$ has 9 phenotypes and $\mathcal{C}_2$ has 8 phenotypes. Therefore,
the fraction of missing data is larger under mapping $h_2$ than under $h_1$. We simulate 1000 
observed phenotypes under both mappings using complete-data model (\ref{inbreed-model}) with 
$p_A = 0.3$, $p_B=0.2$, $p_C=0.2$, $p_D=0.3$ nd $f =$ 0, 0.125, 0.25, 0.375, 0.5. 
For each of these 10 simulated data sets, we estimate allele frequencies
$\hat{p}_A$, $\hat{p}_B$, $\hat{p}_C$, and $\hat{p}_D$ using the EM algorithm and assuming that 
$f=0$.
\begin{figure}[!t]
\begin{center}
  \includegraphics[width=\textwidth]{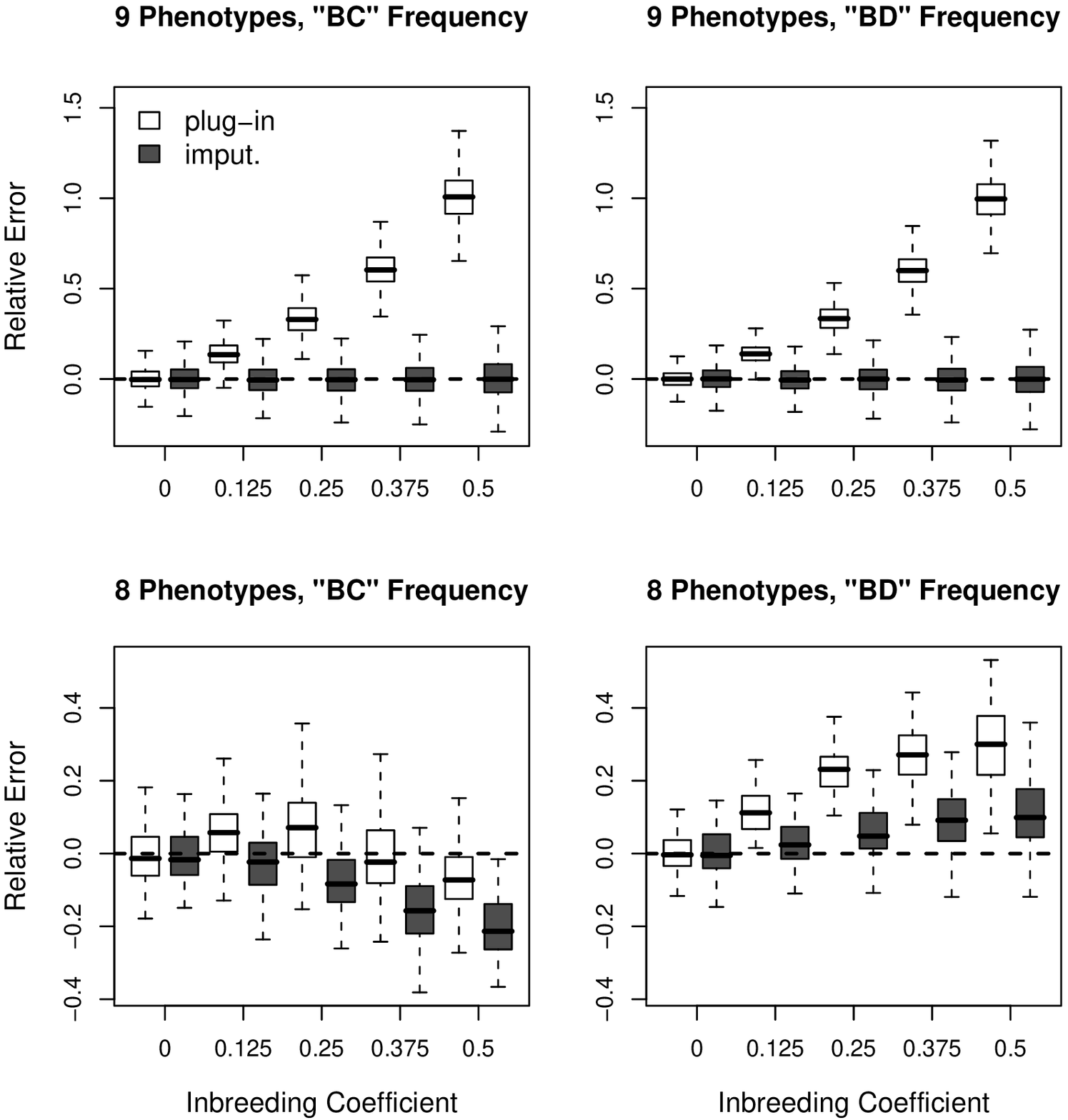}
\end{center}
\vspace{-1cm}
\caption{Genotype frequency estimation. We plot box plots of relative errors, of \conv and \robust estimates of 
genotype frequencies ($\mu_{BC}$ and $\mu_{BD}$) for two incomplete data mappings, with 9 and 8 observed phenotypes.
Each pair of white and grey box plots corresponds to
an inbreeding coefficient that ranges from 0 to 0.5.}
\label{hw-figure}
\end{figure}
\par
For phenotypes that unambiguously correspond to exactly one genotype, the empirical phenotype frequency
can be used to estimate the corresponding genotype frequency. Therefore, it only makes sense to compare 
\conv and \robust estimation for genotypes that correspond to ambiguously defined phenotypes. 
For example, under both $h_1$ and $h_2$  genotypes $(B,C)$ and $(B,D)$ correspond to the phenotype $bcd$.
Suppose our goal is to estimate these genotype frequencies:
$\mu_{BC} = \prob{\mathbf{x}_1=(B,C)} \quad \text{ and } \quad \mu_{BD} = \prob{\mathbf{x}_1=(B,D)}$.
\Conv estimates of these population-level quantities are
\[
\hat{\mu}_{BC}^{pi} = 2 \hat{p}_B \hat{p}_C \quad \text{ and } \quad  \hat{\mu}_{BD}^{pi} = 2 \hat{p}_B \hat{p}_D.
\]
\Robust estimates are obtained as
\[
\hat{\mu}_{BC}^{im} = \frac{n_{bcd}}{n}\frac{\hat{p}_B \hat{p}_C}{\hat{p}_B\hat{p}_C+\hat{p}_B\hat{p}_D}
\quad \text{ and } \quad 
\hat{\mu}_{BD}^{im} = \frac{n_{bcd}}{n}\frac{\hat{p}_B \hat{p}_D}{\hat{p}_B\hat{p}_C+\hat{p}_B\hat{p}_D},
\]
where $n_{bcd} = \sum_{i=1}^n \indfun{y_i = bcd}$ and $n=1000$. Figure~\ref{hw-figure} shows box plots of 
relative errors of \conv and \robust estimators, obtained by repeating the above simulation and estimation 
steps 1000 times.
In the case of 9 phenotypes, corresponding to $h_1$ mapping, the \robust estimation offers remarkable protection
against model misspecification. Decreasing the number of observed phenotypes from 9 to 8 
results in the \robust estimators outperforming the \conv one only for $f=0.125$ 
and $f = 0.25$. 
For the rest of inbreeding coefficient values, \conv estimation produces better estimates of $\mu_{BC}$, 
while \robust estimation offers better estimates of $\mu_{BD}$. However, overall \robust relative errors are still smaller  
than \conv errors.

\section{Labeled evolutionary distances}
\Robust estimation was proposed by \citet{OBrien2009}  in the context of estimation of 
evolutionary distances  between molecular sequences, a standard problem in molecular 
evolutionary biology \citep{Gu1998,ZihengBook}.
Consider a $2\times n$ matrix $\obsdata = \{\obsdatascalar_{ij}\}$, where each $\obsdatascalar_{ij}$ 
takes values in the $\statespace=\{1,\dots,s\}$. 
We assume that all columns in $\obsdata$ are independently generated 
by the same reversible and irreducible continuous-time Markov chain (CTMC) 
$\{\markovchain{t}\}$, defined on the finite state-space $\statespace$ by infinitesimal generator 
$\infgen(\truepar)$. 
This Markov process models the evolution of DNA sequences so that the state space 
$\statespace$ usually consists of 4 nucleotide bases, however, a couple of alternative state-spaces
are also often used. 
Each column $\obsdata_i$ in $\obsdata$ is produced by first drawing $\obsdatascalar_{1i}$ 
from the stationary distribution of $\{X_t\}$, 
$\statdist(\truepar) = (\statdistcomp_1(\truepar),\dots\statdistcomp_s(\truepar))$, 
running the chain for an unknown time $t$ and setting $\obsdatascalar_{2i} = \markovchain{t}$. 
For each realization $i$, we observe only the starting and ending states of the 
Markov chain on the time interval $[0,t]$. Here, model misspecification usually manifests itself
through an incorrect parameterization of the infinitesimal generator, $\infgen(\falsepar)$.
The misspecified likelihood of the observed data is
$\falsesampl(\obsdata;\falsepar) = \prod_{i=1}^n \statdistcomp_{y_{1i}}(\truepar)
  \transprobscalar_{\obsdatascalar_{1i}\obsdatascalar_{2i}}\left(\falsepar,t\right)$, 
where $\transprob(\falsepar,t) = e^{\infgen(\falsepar)t} = \{\transprobscalar_{ij}(\falsepar,t)\}$ 
and  $\transprobscalar_{ij}(\falsepar,t)=\cprob{\markovchain{t}=j}{\markovchain{0}=i}$ are 
finite-time transition probabilities of $\{X_t\}$. Notice that transition probabilities depend on 
$\infgen$ and $t$ only through their product. Therefore, we require the identifiability constraint
$t=1$.
\par
In this example, complete data consist of the full Markov chain trajectory 
$\compdata_i = \{\markovchain{ri}: 0<r<t\}$. A complete-data summary of scientific interest is
$\compsumscalar(\compdata_{1}) = N_{\labelset}$,
the number of transitions of $\markovchain{t}$ during the time interval $[0,1]$, 
labeled by the set of ordered state pairs $\mathcal{L}$. In the absence of complete Markov trajectories,
we are interested in the mean number of labeled transitions of the stationary Markov chain, available
analytically via
\begin{equation} 
  \summeanscalar = \text{E}_{\truepar}\left[\compsumscalar(\compdata_{1})\right] = 
  \text{E}_{\truepar}\left(N_{\labelset}\right) = \statdist(\truepar)^T \infgen_{\labelset}(\truepar)\mathbf{1},
  \label{markov_stat_mean}
\end{equation} 
where $\mathbf{1}$ is an s-dimensional column vectors of 1s and 
$\infgen_{\labelset} = \{\lambda_{uv}\times 1_{\{(u,v) \in \labelset\}}\}$. In molecular evolution, 
this expected number of labeled Markov transitions translates into mean number of labeled mutations,
allowing evolutionary biologists to measure molecular sequence similarity in a flexible manner 
\citep{OBrien2009}.
\par 
The \conv approach for estimating $\mu$ proceeds by first fitting a possibly misspecified 
Markov model, $\infgen(\falsepar)$ and then using the resulting parameter estimates to compute complete-data 
summary expectations. 
More specifically, we obtain $\falseparest = \argmax_{\falsepar} p(\obsdata;\falsepar)$
and obtaining \conv and \robust estimators
\begin{equation*} 
  \hat{\summeanscalar}^{pi} = \statdist(\falseparest)^T \infgen(\falseparest)\mathbf{1}
  \qquad \text{ and } \qquad
  \hat{\summeanscalar}^{im} = 
  \frac{1}{n}\sum_{i=1}^n \text{E}_{\falseparest}\left[N^{\labelset}_1\,|\,\markovchain{0}=\obsdatascalar_{1i},
    \markovchain{1}=\obsdatascalar_{2i}\right].
\end{equation*}  
\citet{OBrien2009} execute two extensive simulation studies that demonstrate that the
\robust estimator offers remarkable protection against misspecification of
a Markovian mutational model. 

\subsection{Complete-data likelihood}
After falsely 
assuming a misspecified model parameterization $\infgen(\falsepar)$ (and $\statdist(\falsepar)$ as a 
result) we condition on the initial Markov chain states and write the misspecified conditional 
complete-data likelihood
\begin{equation}
  \falsesampl(\markovchain{[0,1]};\falsepar) \propto \left[\prod_{u \ne v} 
  \infgencomp_{uv}^{n_{uv}}(\falsepar)\right]\times
  e^{\sum_{u=1}^s T_u \lambda_{uu}(\falsepar)},
  \label{ctmc_comp_like}
\end{equation}
where $n_{uv}$ is the number of times $\markovchain{t}$ instantaneously jumped from $u$ to $v$ and 
$T_u$ is the total time $\markovchain{t}$ spent in state $u$ during the time interval $[0,1]$, both
summed over all $n$ realizations of the Markov chain \citep{Guttorp1995}. 
The complete-data likelihood belongs to the curved exponential family with sufficient statistics
$\mathbf{n} = \{n_{uv}\}_{u \ne v}$ and $\mathbf{T} = (T_1,\dots,T_s)$. 
\par
Nearly all Markov infinitesimal generators used in molecular evolutionary biology fall into the set
$\mathcal{A} = \{\infgen=\{\infgencomp_{uv}\}: \infgencomp_{uv} = \statdistcomp_v \alpha_{uv} 
\text{ for } u \ne v\}$,
where $\boldsymbol{\alpha} = \{\alpha_{uv}\}$ is a symmetric matrix. Such parameterization ensures 
reversibility of the Markov chain, a common assumption in the field of molecular evolution 
\citep{ZihengBook}. 

\subsection{Group-based models}
Notice that the likelihood (\ref{ctmc_comp_like}) simplifies significantly if we assume a reversible
model with equal diagonal entries of $\infgen$:
\begin{equation}
  \falsesampl(\markovchain{[0,1]};\boldsymbol{\alpha}) \propto 
    \prod_{u < v} 
    \alpha_{uv}^{n_{uv}+n_{vu}},
\end{equation}
because $\sum_{u=1}^s T_u = 1$ is the length of the observational time interval.
It turns out that in molecular evolution, only so called group-based models satisfy these 
properties \citep{Evans1993}. 
Group-based Markov  evolutionary models can be defined as continuous-time random walks on 
Abelian groups. If we define an Abelian group on a Markov chain state space $\statespace$ with 
algebraic operation ``$+$'', then entries of the corresponding group-based CTMC generator 
$\infgen$ must satisfy $\infgencomp_{uv} = g(u-v)$ for some function 
$g: \statespace \rightarrow [0,\infty)$. For example, the most general group-based model on the
state space of DNA bases $\{A,G,C,T\}$ is a Kimura three-parameter model with
\begin{equation}
  \infgen^{\text{K3P}}(\alpha,\beta,\gamma) = 
  \begin{pmatrix}
    - & \alpha & \beta & \gamma\\
    \alpha & - & \gamma & \beta\\
    \beta & \gamma & - & \alpha\\
    \gamma & \beta & \alpha & -
  \end{pmatrix},
  \label{k3st}
\end{equation}
corresponding to the Klein group $\mathbb{Z}_2\oplus\mathbb{Z}_2$ \citep{Evans1993}.
\par
Group-based models, constructed with algebraic symmetry in mind, find extensive use in 
statistical phylogenetics \citep{Sturmfels2005,Steel1998}.
For us, these models are appealing because they turn the completed-data CTMC likelihood into
the regular exponential family form. If we break all possible DNA mutations into three classes and
define their corresponding counts 
$N_{AG,CT} = n_{AG}+n_{GA}+n_{CT}+n_{TC}$,
$N_{AC,GT} = n_{AC}+n_{CA}+n_{GT}+n_{TG}$, and
$N_{AT,GC} = n_{AT}+n_{TA}+n_{GC}+n_{CG}$,
then these counts form the sufficient statistics for the Kimura three-parameter model. From our discussion
of the regular exponential family it follows that \conv and \robust estimates of 
$\text{E}_{\alpha,\beta,\gamma}\left(c_1N_{AG,CT}+c_2 N_{AC,GT}+c_3N_{AT,GC}\right)$ will coincide exactly 
regardless of the true sampling model and of the choice of constants $c_1$, $c_2$, and $c_3$. This 
fact was not noticed by \citet{OBrien2009}, because the authors did not consider group-based models
explicitly in their work.

\subsection{A closer look at observed data likelihood equations}
Instead of invoking properties of the regular exponential family, one can find more general
conditions under which \robust and \conv estimates of labeled evolutionary distances coincide,
as demonstrated by the theorem below.
\begin{theorem}
  \label{ctmc_like_thm}
  Let $\obsdata = \{\obsdatascalar_{ij}\}, i=1,2, j=1,\dots,n$, be a pairwise sequence alignment generated
  by a CTMC with an unknown infinitesimal generator $\infgen(\truepar)$ as described at the beginning of
  this section. We take $\infgen(\falsepar)$ to be a misspecified model and 
  $\falseparest = (\falseparestcomp{1},\dots,\falseparestcomp{r})$ to 
  be the corresponding maximum likelihood estimator obtained from the observed data $\obsdata$. If 
  \begin{equation} 
    \infgen_{\mathcal{L}}(\falseparest) 
    - \mathbf{I} \times \statdist^T(\falseparest) \infgen_{\mathcal{L}}(\falseparest) \mathbf{1}
    \in 
    \left<\frac{\partial \infgen(\falsepar)}{\partial \falseparcomp{1}}\Big|_{\falsepar=\falseparest}, \dots, 
      \frac{\partial \infgen(\falsepar)}{\partial \falseparcomp{d}}\Big|_{\falsepar=\falseparest} 
    \right>,
    \label{thm_cond}
  \end{equation}
where $\mathcal{L} \subset S^2\setminus \{(i,i): i \in S\}$ is a set of ordered Markov state pairs and 
$\infgen_{\labelset} = \{\infgencomp_{uv}\times 1_{\{(u,v) \in \labelset\}}\}$, then
\begin{equation*} 
 \text{E}_{\falseparest}\left(N_{\labelset} \right) =
  \frac{1}{n}\sum_{i=1}^n \text{E}_{\falseparest}\left(N_{\labelset}\,|\,\markovchain{0}=\obsdatascalar_{1i},
    \markovchain{1}=\obsdatascalar_{2i}\right),  
\end{equation*}
where $N_{\labelset}$ is the unobserved number of Markov chain transitions labeled by the set $\labelset$. 
\label{der_thm}
\end{theorem}
To illustrate the above theorem, consider a Kimura two-parameter model 
$\infgen^{\text{K2P}}(\alpha,\beta) = \infgen^{\text{K3P}}(\alpha,\beta,\beta)$, obtained
by setting $\gamma = \beta$ in matrix (\ref{k3st}) \citep{Kimura1980}. For both of these models, 
the stationary distribution is $\statdist^T = (0.25,0.25,0.25,0.25)$. Let 
\begin{align}
  \label{transitions}
  \labelset_1 &= \{(A,G),(G,A),(C,T),(T,C)\} \text{ and }\\
  \label{transversions}
  \labelset_2 &= \{(A,C),(C,A),(A,T),(T,A),(C,G),(G,C),(T,G),(G,T)\}
\end{align}
be two mutational classes of interest. The partial derivatives of the Kimura two-parameter generator,
\begin{align*}
    \frac{\partial}{\partial \alpha} \infgen^{\text{K2P}}(\alpha,\beta) &= 
    \frac{1}{\alpha}\left[\infgen_{\labelset_1} - \mathbf{I} \times 
      \statdist^t \infgen_{\labelset_1}\mathbf{1}\right] \text{ and}\\
    \frac{\partial}{\partial \beta} \infgen^{\text{K2P}}(\alpha,\beta) &= 
    \frac{1}{\beta}\left[\infgen_{\labelset_1} - \mathbf{I} \times 
      \statdist^t \infgen_{\labelset_2}\mathbf{1}\right],
\end{align*}
satisfy condition (\ref{thm_cond}). Therefore, Theorem 1 says that \conv and \robust estimators of 
$\text{E}_{\alpha,\beta}\left(N_{\labelset_1}\right)$ and $\text{E}_{\alpha,\beta}\left(N_{\labelset_2}\right)$
coincide exactly. Of course this example reiterates the fact that
complete-data likelihood of the Kimura two-parameter model belongs to the regular exponential family
with sufficient statistics $N_{\labelset_1}$ and $N_{\labelset_2}$.

\subsection{Misspecified Kimura model: asymptotic behavior}
Studying asymptotic properties of our \robust estimator is challenging in general even for the 
specific problem of the evolutionary distance estimation. Therefore, we turn to an elementary example to 
obtain some basic asymptotic results. First, we introduce the simplest group-based model on the nucleotide
state space, known as a Jukes-Cantor model. The infinitesimal generator of this Markov chain is obtained
by setting $\alpha=\beta$ in the Kimura two-parameter model, 
$\infgen^{\text{JC}}(\gamma)=\infgen^{\text{K2P}}(\gamma,\gamma)$ \citep{Jukes1969}. 
\begin{theorem}
  \label{limit_thm}
  Assume that observed sequence data $\obsdata$ was generated from the Kimura two-parameter model with 
  generator $\infgen^{\text{K2P}}(\alpha,\beta)$. Let $\hat{\gamma}$ be the maximum likelihood estimate,
  obtained by fitting a Jukes-Cantor model with generator 
  $\infgen^{\text{JC}}(\gamma)$ to $\obsdata$. Then as the number of columns in $\obsdata$, $n$, approaches
  infinity,
  \begin{align*}
    \text{E}_{\hat{\gamma}}\left(N_{\labelset_1} \right) &\asarrow \beta - \frac{1}{4}\ln\left(\frac{1+2e^{2(\beta-\alpha)}}{3}\right)\\
    \begin{split}
      \frac{1}{n}\sum_{i=1}^n \text{E}_{\hat{\gamma}}\left(N_{\labelset_1}\,|\,\markovchain{0}=\obsdatascalar_{1i},
        \markovchain{1}=\obsdatascalar_{2i}\right) &\asarrow
      \left[\beta - \frac{1}{4}\ln\left(\frac{1+2e^{2(\beta-\alpha)}}{3}\right)\right]\\
      &\times 
      \left[1+\frac{4(e^{-4\beta} + 2e^{-2(\alpha+\beta)})(e^{-4\beta}-e^{-2(\alpha+\beta)})}
        {3(3-e^{-4\beta}-e^{-2(\alpha+\beta)})}\right],
      \end{split}
    \end{align*}
  where $\labelset_1$ is defined by equation (\ref{transitions}).
\end{theorem}
\begin{figure}[!t]
\begin{center}
  \includegraphics[width=\textwidth]{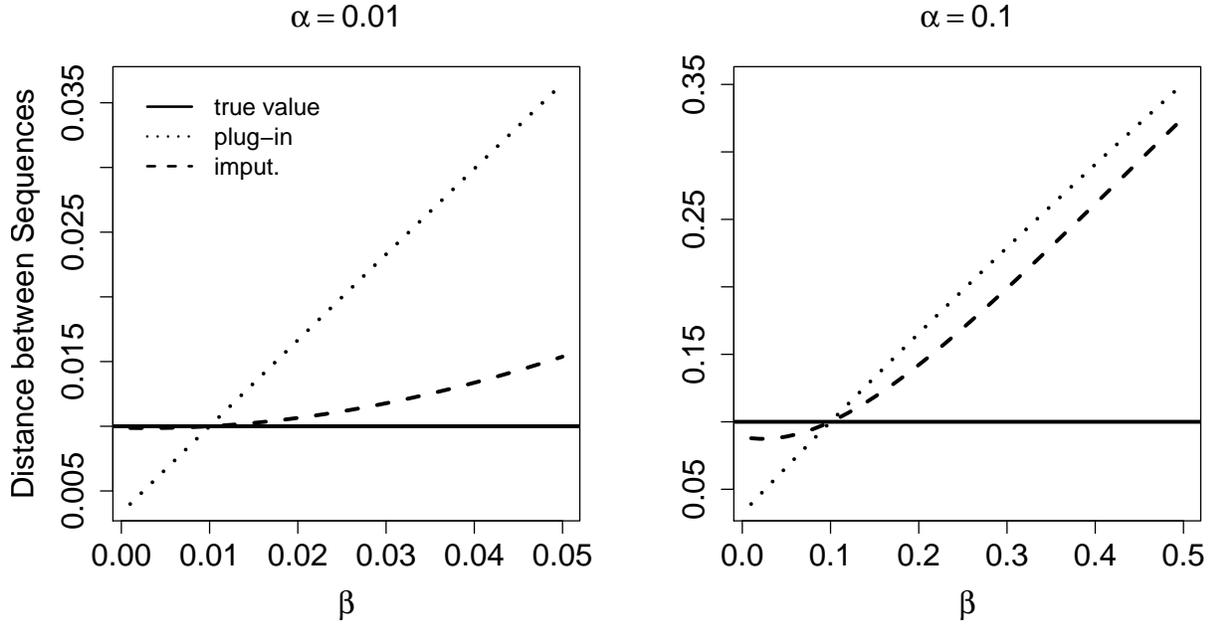}
\end{center}
\vspace{-1.5cm}
\caption{\small Estimator limits in the Kimura model. In both panels of the figure we plot the true value of the 
  mean number of labeled mutations $\mu=\text{E}\left(N_{\labelset}\right)$ (solid line), the a.s. limits
  of the \conv (dotted line) and \robust (dashed line) estimators.} 
\label{jc_kp2}
\end{figure}
\begin{corollary}
  \label{limit_cor}
  Under the conditions of Theorem \ref{limit_thm}, define
  \begin{align*}
    \mu = \text{E}_{\alpha,\beta}\left(N_{\labelset_1}\right), \qquad
    \mu^{pi}_{\infty} = \lim_{n \rightarrow \infty} \text{E}_{\hat{\gamma}}\left(N_{\labelset_1} \right), \qquad
    \mu^{im}_{\infty} = \lim_{n \rightarrow \infty}  
    \frac{1}{n}\sum_{i=1}^n \text{E}_{\hat{\gamma}}\left(N_{\labelset_1}\,|\,\markovchain{0}=\obsdatascalar_{1i},
      \markovchain{1}=\obsdatascalar_{2i}\right).
  \end{align*}
  Then $|\mu^{im}_{\infty} - \mu| < |\mu^{pi}_{\infty} - \mu|$ when $\alpha \ne \beta$. 
  In other words, the \robust estimator asymptotically is always better than the \conv one. 
\end{corollary}
\par
To illustrate the above theorem and its corollary we plot the true value ($\mu$) and a.s. limits of the 
\conv ($\mu^{pi}_{\infty}$) and \robust ($\mu^{im}_{\infty}$) estimators as function of $\beta$ 
in Figure~\ref{jc_kp2}. We fix $\alpha=0.01$ in the left panel and $\alpha=0.1$ in the right panel. 
Roughly speaking, the left panel shows the behavior of the estimators when the overall mutation rate is low,
 while the right panel corresponds to a high mutation rate scenario. The lower the mutation rate, the better 
our \robust estimator behaves asymptotically. This property of the \robust estimation is expected, because
low mutation rate translates into lower fraction of missing data, which in turn makes the 
\robust estimation more powerful. We have already seen this behavior of the \robust estimator 
in the previous examples.


\section{Bayesian implementation}
\subsection{General recipe}
Although all examples so far were analyzed from the maximum likelihood perspective, 
one can easily perform imputation-based estimation in a Bayesian framework. To accomplish this, we first need 
to assign a prior distribution $p(\falsepar)$  to the parameters of our misspecified model 
$\falsesampl(\obsdata;\falsepar)$. We assume that it is possible to obtain either the posterior distribution
$\falsesampl(\falsepar\,|\,\obsdata)$ or
the augmented posterior $\falsesampl(\falsepar,\compdata \,|\,\obsdata)$, 
possibly approximating these distributions via Markov chain Monte Carlo (MCMC) \citep{Tanner1987}. 
Using these posterior 
distributions, we define \conv and two \robust predictive distributions
\begin{equation*}
  p\left(\text{E}_{\falsepar}[\compsum(\compdata_1)]\,\big|\,\obsdata\right), \qquad
  p\left(\frac{1}{n}\sum_{i=1}^n\compsum(\compdata_i)\,\Big|\,\obsdata\right), \qquad \text{ and } \qquad
  p\left(\frac{1}{n}\sum_{i=1}^n \text{E}_{\falsepar}[\compsum(\compdata_i)\,|\,
    \obsdata_i]\,\Big|\,\obsdata\right).
\end{equation*}
As before, we hope that the latter two will provide us some protection against model misspecification.
These last two predictive distributions have the same mean, but conditioning reduces the variance of the
third distribution. This is similar to Rao-Blackwellization in Monte Carlo sampling \citep{Casella1996}, but  
since we are working under the assumption of model misspecification,
smaller variance is not necessary a desirable property of a predictive distribution.
\subsection{Bayesian estimation of genotype frequencies}
To illustrate the Bayesian implementation of our procedure, we revisit the genotype frequency estimation example.
We generate 10 phenotype samples using the two genotype-to-phenotype mappings defined in
Table~\ref{hw-maps} and setting the inbreeding coefficient $f$ and true allele frequencies to the values
we used in the original example. We place $\text{Dirichlet}(1,1,1,1)$ prior on allele frequencies and approximate
the posterior distribution of complete data (genotype counts) and allele frequencies via Gibbs sampling.
\par
\begin{figure}[!t]
\begin{center}
  \includegraphics[width=\textwidth]{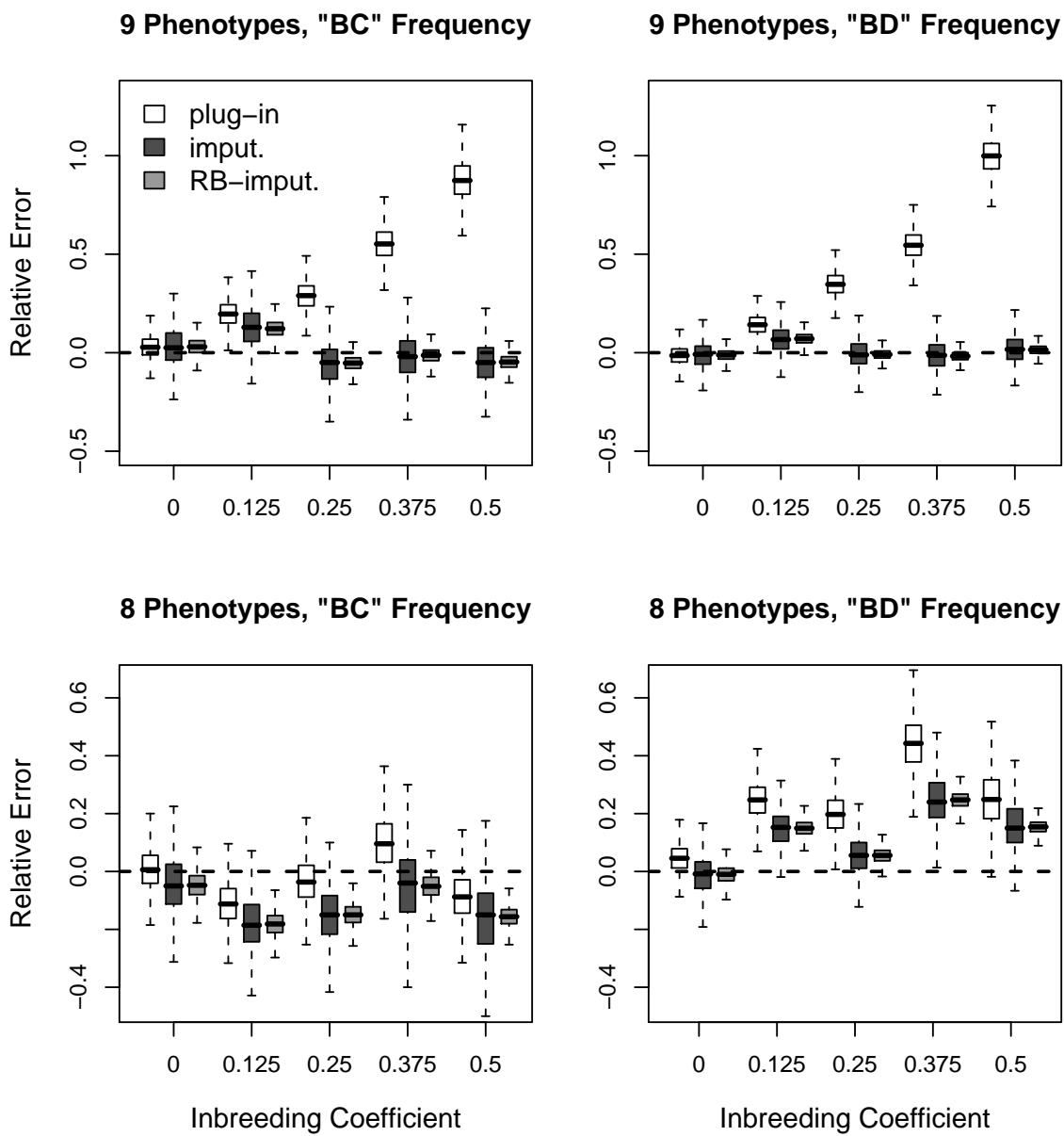}
\end{center}
\vspace{-1cm}
\caption{Genotype frequency estimation. We plot box plots of relative errors of plug-in, imputation, and 
Rao-Blackwellized \robust estimates of 
genotype frequencies ($\mu_{BC}$ and $\mu_{BD}$) for two incomplete data mappings, with 9 and 8 observed phenotypes.
Each trio of white, dark grey, and light grey box plots corresponds to
an inbreeding coefficient that ranges from 0 to 0.5.}
\label{hw-bayes}
\end{figure}

Recall that our goal is to estimate genotype frequency $\mu_{k,l} = \prob{\mathbf{x}_1=(g_k,g_l)}$. For
$(g_k,g_l) = (B,C)$ and $(g_k,g_l) = (B,D)$, we report posterior distributions of
\[
2 p_k p_l, \qquad \frac{1}{n}\sum_{i=1}^n \indfun{\compdata_i = (g_k,g_l)} = \frac{m_{kl}}{n}, \quad \text{ and } \quad 
  \frac{1}{n}\sum_{i=1}^n \cexpect{\compdata_i = (g_k,g_l)}{\obsdatascalar_i}{(}{)} = 
\frac{n_j 2 p_k p_l}{n(2 p_B p_C + 2 p_B p_D)},
\]
where $m_{kl} = \sum_{i=1}^n \indfun{\compdata_i = (g_k,g_l)}$ and 
$n_j = \sum_{i=1}^n \indfun{\obsdatascalar_i = BCD}$.  We report box plots of these posterior distributions
in Figure~\ref{hw-bayes}. These box plots are not directly comparable to results in Figure~\ref{hw-figure}, 
because our Bayesian analysis is based only on ten data sets, while the maximum analysis was done on 10,000 
simulated data sets, one thousand for each value of $f$ and for each genotype-to-phenotype mapping.
To make these analyses comparable, one can study frequentist properties of Bayesian \conv and \robust 
estimators based, for example, on the posterior median-based estimators of allele frequencies and genotype
counts. Our Bayesian results are nonetheless consistent with the maximum likelihood analysis: \robust
estimators outperform the \conv estimate in the case of nine phenotypes, none of the estimators have
a uniform advantage across all values of the inbreeding coefficient in the eight phenotype case.
\par
We end our discussion of the Bayesian implementation of our \robust estimation by pointing out that this 
inferential framework is already being used in evolutionary biology, albeit somewhat informally 
\citep{Zhai2007, Minin2008b}. These methods extend the idea of \robust evolutionary distance estimation 
to multiple sequences. 

\section{Discussion}
We generalize the notion of imputation estimators and demonstrate that such estimators can be useful in a variety
of incomplete data problems under model misspecification. We use simulations as our main tool in the first 
two examples and provide some simple asymptotic results in our last example. So far, our experience suggests
that imputation estimators perform very well under mild model misspecification and when the fraction of 
missing data is reasonably small. 
Intuitively, it is clear that imputation estimators should be more successful as the amount of 
missing data decreases, because in the absence of missing information these estimators turn into sample means, 
which are model-free and consistent estimates of appropriate population-level quantities. However, to make this
intuition useful, we need to connect formally efficiency of imputation estimators with the amount of missing data 
and degree of model misspecification. We hope to be able to make these connections in our future work. 
\par  
Studying sampling properties of \robust estimators proved to be difficult in general, 
especially since in practice the true sampling density of the observed data is unknown. In fact, in all our examples,
we do not discuss how to compute the variance of the maximum likelihood-based \robust estimators. 
We recommend to use nonparametric bootstrap to explore sampling properties of \robust estimators. 
However, one should interpret bootstrap results with care, because \robust 
estimators remain biased even asymptotically. Similar care needs to be applied to the interpretation of 
predictive distributions in the Bayesian context. 
\par
\par
Although we have not emphasized this throughout the paper, \robust estimators are usually easy to compute, which
makes them particularly useful when a compromise between model complexity and computational efficiency results
in an intentionally misspecified model. In our examples of model misspecification, we considered 
Gaussian mixture components, Hardy-Weinberg genotype frequencies, and parametric Markov models of 
DNA mutation. All these highly popular models owe a large portion of their success to their computational 
tractability. We argue
that imputation estimators can take these and many other simple and computationally efficient models 
one step further outside of their usual domain of application.

\section*{Acknowledgments}
VNM was partially supported by the UW Royalty Research Fund and by the National Scientific Foundation grant 
No.\ DMS-0856099. AS was partially supported by the National Scientific Foundation grant No.\ DMS-0804587. 
JDO was partially supported by the Wellcome Trust grant No.\ WT082930MA.
We thank Adrian Raftery and his working group for their comments on model-based clustering. 


\bibliography{list_of_references}

\begin{thebibliography}{32}
\providecommand{\natexlab}[1]{#1}
\providecommand{\url}[1]{\texttt{#1}}
\expandafter\ifx\csname urlstyle\endcsname\relax
  \providecommand{\doi}[1]{doi: #1}\else
  \providecommand{\doi}{doi: \begingroup \urlstyle{rm}\Url}\fi

\bibitem[Allen and Satten(2008)]{Allen2008}
A.S. Allen and G.A. Satten.
\newblock Robust estimation and testing of haplotype effects in case-control
  studies.
\newblock \emph{Genetic Epidemiology}, 32:\penalty0 29--40, 2008.

\bibitem[Ball and Milne(2005)]{Ball2005}
F.~Ball and R.K. Milne.
\newblock Simple derivations of properties of counting processes associated
  with {M}arkov renewal processes.
\newblock \emph{Journal of Applied Probability}, 42:\penalty0 1031--1043, 2005.

\bibitem[Casella and Robert(1996)]{Casella1996}
G.~Casella and C.~Robert.
\newblock Rao-{B}lackwellisation of sampling schemes.
\newblock \emph{Biometrika}, 83:\penalty0 81--94, 1996.

\bibitem[Ceppelini et~al.(1955)Ceppelini, Siniscalco, and Smith]{Ceppelini1955}
R.~Ceppelini, M.~Siniscalco, and C.A.B. Smith.
\newblock The estimation of gene frequencies in a random mating population.
\newblock \emph{Annals of Human Genetics}, 20:\penalty0 97--115, 1955.

\bibitem[Chen et~al.(2009)Chen, Chatterjee, and Carroll]{Chen2009}
Y.-H. Chen, N.~Chatterjee, and R.J. Carroll.
\newblock Shrinkage estimators for robust and efficient inference in
  haplotype-based case-control studies.
\newblock \emph{Journal of the American Statistical Association}, 104:\penalty0
  220--233, 2009.

\bibitem[Dempster et~al.(1977)Dempster, Laird, and Rubin]{Dempster1977}
A.P. Dempster, N.M. Laird, and D.B. Rubin.
\newblock Maximum likelihood from incomplete data via the {EM} algorithm.
\newblock \emph{Journal of the Royal Statistical Society, Series B},
  39:\penalty0 1--38, 1977.

\bibitem[Evans and Speed(1993)]{Evans1993}
S.N. Evans and T.P. Speed.
\newblock Invariants of some probability models used in phylogenetic inference.
\newblock \emph{The Annals of Statistics}, 21:\penalty0 355--377, 1993.

\bibitem[Fraley and Raftery(2002)]{Fraley2002}
C.~Fraley and A.E. Raftery.
\newblock Model-based clustering, discriminant analysis, and density
  estimation.
\newblock \emph{Journal of the American Statistical Association}, 97:\penalty0
  611--631, 2002.

\bibitem[Fraley and Raftery(2003)]{Fraley2003}
C.~Fraley and A.E. Raftery.
\newblock Enhanced software for model-based clustering, density estimation, and
  discriminant analysis: Mclust.
\newblock \emph{Journal of Classification}, 20:\penalty0 263--286, 2003.

\bibitem[Gu and Li(1998)]{Gu1998}
X.~Gu and W.H. Li.
\newblock Estimation of evolutionary distances under stationary and
  nonstationary models of nucleotide substitution.
\newblock \emph{Proceedings of the National Academy of Sciences, USA},
  95:\penalty0 5899--5905, 1998.

\bibitem[Guttorp(1995)]{Guttorp1995}
P.~Guttorp.
\newblock \emph{Stochastic Modeling of Scientific Data}.
\newblock Chapman \& Hall, Suffolk, Great Britain, 1995.

\bibitem[Hardy(1908)]{Hardy1908}
G.H. Hardy.
\newblock Mendelian proportions in a mixed population.
\newblock \emph{Science}, 28:\penalty0 49--50, 1908.

\bibitem[Jukes and Cantor(1969)]{Jukes1969}
T.H Jukes and C.R. Cantor.
\newblock \emph{Evolution of protein molecules}, pages 21--32.
\newblock Academic Press, New York, 1969.

\bibitem[Kang and Schafer(2007)]{Kang2007}
J.D.Y. Kang and J.L. Schafer.
\newblock Demystifying double robustness: a comparison of alternative
  strategies for estimating a population mean from incomplete data.
\newblock \emph{Statistical Science}, 22:\penalty0 523--539, 2007.

\bibitem[Kimura(1980)]{Kimura1980}
M.~Kimura.
\newblock A simple method for estimating evolutionary rates of base
  substitutions through comparative studies of nucleotide sequences.
\newblock \emph{Journal of Molecular Evolution}, 16:\penalty0 111--120, 1980.

\bibitem[Kraft et~al.(2005)Kraft, Cox, Paynter, Hunter, and Vivo]{Kraft2005}
P.~Kraft, D.G. Cox, R.A. Paynter, D.~Hunter, and I.~De Vivo.
\newblock Accounting for haplotype uncertainty in matched association studies:
  a comparison of simple and flexible techniques.
\newblock \emph{Genetic Epidemiology}, 28:\penalty0 261--272, 2005.

\bibitem[Little and An(2004)]{Little2004}
R.~Little and H.~An.
\newblock Robust likelihood-based analysis of multivariate data with missing
  values.
\newblock \emph{Statistica Sinica}, 14:\penalty0 949--968, 2004.

\bibitem[Minin and Suchard(2008{\natexlab{a}})]{Minin2008a}
V.N. Minin and M.A. Suchard.
\newblock Counting labeled transitions in continuous-time {M}arkov models of
  evolution.
\newblock \emph{Journal of Mathematical Biology}, 56:\penalty0 391--412,
  2008{\natexlab{a}}.

\bibitem[Minin and Suchard(2008{\natexlab{b}})]{Minin2008b}
V.N. Minin and M.A. Suchard.
\newblock Fast, accurate and simulation-free stochastic mapping of discrete
  traits.
\newblock \emph{Philosophical Transactions of the Royal Society B: Biological
  Sciences}, 363:\penalty0 3985--3995, 2008{\natexlab{b}}.

\bibitem[O'Brien et~al.(2009)O'Brien, Minin, and Suchard]{OBrien2009}
J.D. O'Brien, V.N. Minin, and M.A. Suchard.
\newblock Learning to count: Robust estimates for labeled distances between
  molecular sequences.
\newblock \emph{Molecular Biology and Evolution}, 26:\penalty0 801--814, 2009.

\bibitem[Redner and Walker(1984)]{Redner1984}
R.A. Redner and H.F. Walker.
\newblock Mixture densities, maximum likelihood and the {EM} algorithm.
\newblock \emph{SIAM Review}, 26:\penalty0 195--239, 1984.

\bibitem[Rosenbaum and Rubin(1983)]{Rosenbaum1983}
P.R. Rosenbaum and D.B. Rubin.
\newblock The central role of the propensity score in observational studies for
  causal effects.
\newblock \emph{Biometrika}, 70:\penalty0 41--55, 1983.

\bibitem[Steel et~al.(1998)Steel, Hendy, and Penny]{Steel1998}
M.~Steel, M.D. Hendy, and D.~Penny.
\newblock Reconstructing phylogenies from nucleotide pattern probabilities: A
  survey and some new results.
\newblock \emph{Discrete Applied Mathematics}, 88:\penalty0 367--396, 1998.

\bibitem[Sturmfels and Sullivant(2005)]{Sturmfels2005}
B.~Sturmfels and S.~Sullivant.
\newblock Toric ideals of phylogenetic invariants.
\newblock \emph{Journal of Computational Biology}, 12:\penalty0 204--228, 2005.

\bibitem[Sundberg(1974)]{Sundberg1974}
R.~Sundberg.
\newblock Maximum likelihood theory for incomplete data from an exponential
  family.
\newblock \emph{Scandinavian Journal of Statistics}, 1:\penalty0 49--58, 1974.

\bibitem[Tanner and Wong(1987)]{Tanner1987}
M.A. Tanner and W.H. Wong.
\newblock The calculation of posterior distributions by data augmentation.
\newblock \emph{Journal of the American Statistical Association}, 82:\penalty0
  528--540, 1987.

\bibitem[Tsiatis and Davidian(2007)]{Tsiatis2007}
A.~Tsiatis and M.~Davidian.
\newblock Comment: Demystifying double robustness: a comparison of alternative
  strategies for estimating a population mean from incomplete data.
\newblock \emph{Statistical Science}, 22:\penalty0 569--573, 2007.

\bibitem[Tsiatis(2006)]{TsiatisBook}
A.A. Tsiatis.
\newblock \emph{Semiparametric Theory and Missing Data}.
\newblock Springer, New York, 2006.

\bibitem[van~der Vaart and Wellner(2000)]{vanderVaartWellnerBook}
A.W. van~der Vaart and J.A. Wellner.
\newblock \emph{Weak convergence and empirical processes}.
\newblock Springer-Verlag, New York, corrected second printing edition, 2000.

\bibitem[Weinberg(1908)]{Weinberg1908}
W.~Weinberg.
\newblock \"{U}ber den nachweis der vererbung beim menschen.
\newblock \emph{Jahreshefte des Vereins f\"{u}r vaterl\"{a}ndische Naturkunde
  in W\"{u}ttemberg}, 64:\penalty0 368--382, 1908.

\bibitem[Yang(2006)]{ZihengBook}
Z.~Yang.
\newblock \emph{Computational Molecular Evolution}.
\newblock Oxford University Press, USA, 2006.

\bibitem[Zhai et~al.(2007)Zhai, Slatkin, and Nielsen]{Zhai2007}
W.~Zhai, M.~Slatkin, and R.~Nielsen.
\newblock Exploring variation in the {$d_N/d_S$} ratio among sites and lineages
  using mutational mappings: applications to the influenza virus.
\newblock \emph{Journal of Molecular Evolution}, 65:\penalty0 340--348, 2007.

\end{thebibliography}

\clearpage

\renewcommand{\theequation}{A-\arabic{equation}}
\setcounter{equation}{0}
\section*{Appendix}

\begin{proof}[Proof of Theorem \ref{ctmc_like_thm}]
Defining $m_{kl} = \sum_{i=1}^n 1_{\{\obsdatascalar_{1i}=k,\obsdatascalar_{2i}=l\}}$, the misspecified complete-data 
log-likelihood takes the following form:
\begin{equation}
  \displaystyle l(\obsdata, \falsepar) = 
  \sum_{k\in S} \sum_{l \in S} m_{kl} \ln p_{kl}(\falsepar,1),
  \label{ctmc_log_like}
\end{equation}
where $p_{kl}(\falsepar,1)$ is the probability of $X_1=l$ conditional
on starting $X_0=k$. Recall that $\mathbf{P}(\falsepar,1) = e^{\Lambda(\falsepar)} = \{p_{ij}(\falsepar,1)\}$.
Differentiating (\ref{ctmc_log_like}) with respect model parameters, we arrive at the likelihood equations
\begin{equation}
  \begin{split}
    \sum_{k \in E} \sum_{l \in E} 
    \frac{m_{kl}}{p_{kl}(\falsepar,1)} 
     \frac{\partial p_{kl}(\falsepar,1)}{\partial \falseparcomp{j}} = 0, j = 1,\dots,r.
  \end{split}
  \label{like_eqns}
  \end{equation}
From backward Kolmogorov equation $\frac{\text{d} \mathbf{P}(\falsepar,t)}{\text{d}t} = 
\Lambda(\falsepar)\mathbf{P}(\falsepar,t)$ with initial condition $\mathbf{P}(\falsepar,0)=\mathbf{I}$,
we derive the following integral expression for the partial derivatives of transition probabilities:
\begin{equation*}
  \frac{\partial}{\partial \falseparcomp{j}} \text{P}\left(\falsepar,1\right) = 
  \int_0^1 e^{\boldsymbol{\Lambda}(\falsepar)} \frac{\partial}{\partial \falseparcomp{j}} 
  \boldsymbol{\Lambda}(\falsepar)  e^{\boldsymbol{\Lambda}(\falsepar)(1-\tau)}
  \text{d}\tau.
\end{equation*} 
\par
Next, we write the \robust estimator in terms of $m_{kl}$,
\begin{equation*} 
  \frac{1}{n}\sum_{i=1}^n \text{E}_{\falseparest}\left(N_{\labelset}\,|\,\markovchain{0}=\obsdatascalar_{1i},
    \markovchain{1}=\obsdatascalar_{2i}\right)= 
  \frac{1}{n}
  \sum_{k \in S} \sum_{l \in S} 
  \frac{m_{kl}}{p_{kl}(\falseparest,1)}  
  \text{E}_{\falseparest}\left(N_{\labelset} 1_{\{X_1=l\}}\,|\,X_0=k\right),
\end{equation*}  
where
\begin{equation} 
  \text{E}_{\falseparest}\left(N_{\labelset} 1_{\{X_1=l\}}\,|\,X_0=k\right)=
  \left\{\int_0^1 e^{\infgen(\falseparest)}
    \infgen_{\labelset}(\falseparest)  e^{\infgen(\falseparest)(1-\tau)}
    \text{d}\tau\right\}_{kl}.
  \label{cond_mean}
\end{equation}
Derivation of the formula (\ref{cond_mean}) can be found in \citep{Ball2005} or \citep{Minin2008a}.
Condition (\ref{thm_cond}) says that there exist real constants $c_1,\dots,c_r$ such that 
\begin{equation*}
  \infgen_{\mathcal{L}}(\falseparest) 
  - \mathbf{I} \times \statdist^T(\falseparest) \infgen_{\mathcal{L}}(\falseparest) \mathbf{1}
  = 
  \sum_{i=1}^r c_i \frac{\partial \infgen(\falsepar)}{\partial \falseparcomp{i}}\Big|_{\falsepar=\falseparest}.
\end{equation*}
Therefore, the difference between the \conv and \robust estimators becomes
\begin{equation*} 
  \begin{split}
    &\frac{1}{n}\sum_{i=1}^n \text{E}_{\falseparest}\left(N_{\labelset}\,|\,\markovchain{0}=\obsdatascalar_{1i},
      \markovchain{1}=\obsdatascalar_{2i}\right) - \text{E}_{\falseparest}\left(N_{\labelset} \right) = \\ 
    &\frac{1}{n}\sum_{k \in S} \sum_{l \in S}\frac{m_{kl}}{p_{kl}(\falseparest,1)}
      \left\{\int_0^1 e^{\infgen(\falseparest)} 
        [\infgen_{\labelset}(\falseparest) 
        -\mathbf{I} \times \statdist^T(\falseparest) 
        \infgen_{\labelset}(\falseparest) \mathbf{1}]
        e^{\infgen(\falseparest)(1-\tau)}\text{d}\tau\right\}_{kl} = \\
      &\frac{1}{n}\sum_{i=1}^r c_i 
      \sum_{k \in S} \sum_{l \in S}\frac{m_{kl}}{p_{kl}(\falseparest,1)}
      \left\{\int_0^1 e^{\infgen(\falseparest)} 
      \frac{\partial \infgen(\falsepar)}{\partial \falseparcomp{i}}\Big|_{\falsepar=\falseparest}
        e^{\infgen(\falseparest)(1-\tau)}\text{d}\tau\right\}_{kl} = \\
      &\frac{1}{n}\sum_{i=1}^r c_i 
      \sum_{k \in S} \sum_{l \in S}\frac{m_{kl}}{p_{kl}(\falseparest,1)}
      \frac{\partial p_{kl}\left(\falsepar,1\right)}{\partial \falseparcomp{i}}\Big|_{\falsepar=\falseparest}=0,
  \end{split}
\end{equation*}
because $\falseparest$ satisfies likelihood equations (\ref{like_eqns}).
\end{proof}

\begin{proof}[Proof of Theorem \ref{limit_thm}]
  As before, let $m_{kl} = \sum_{i=1}^n \indfun{y_{1i}=k,y_{2i}=l}$. Using these site counts, define
  \[
  m_{\labelset_1} = \sum_{(k,l)\in \labelset_1} m_{kl},\  m_{\labelset_2} = \sum_{(k,l)\in \labelset_2} m_{kl},\  
  m_D = \sum_{k=l} m_{kl},\ 
  f_{\labelset_1} = \frac{m_{\labelset_1}}{n},\ 
  f_{\labelset_2} = \frac{m_{\labelset_2}}{n},\ 
  f_D = \frac{m_D}{n},
  \]
  where $\labelset_2$ is defined by equation (\ref{transversions}).
  Transition probabilities of the Kimura two-parameter model are obtained as
  \begin{equation}
    p_{kl}(\alpha,\beta,t) =
    \begin{cases}
      \frac{1}{4} + \frac{1}{4}e^{-4\beta t} - \frac{1}{2}e^{-2(\alpha+\beta)t} &\text{ if } (k,l) 
      \in \labelset_1,\\
      \frac{1}{4} - \frac{1}{4}e^{-4 \beta t} &\text{ if } (k,l) \in \labelset_2,\\
      \frac{1}{4} + \frac{1}{4}e^{-4\beta t} + \frac{1}{2}e^{-2(\alpha+\beta)t} &\text{ if } k=l.    
    \end{cases}
    \label{k2p_prob}
  \end{equation}
  Since the stationary distribution of the Kimura two-parameter model is uniform, 
  $(m_{\labelset_1},m_{\labelset_2},m_D) \sim \text{Multinomial}(p_{\labelset_1},p_{\labelset_2},p_D)$,
  where 
  \[
  \begin{split}
    p_{\labelset_1} &= \sum_{(k,l)\in \labelset_1} \frac{1}{4}p_{kl}(\alpha,\beta,1) = 
    \frac{1}{4} + \frac{1}{4}e^{-4\beta} - \frac{1}{2}e^{-2(\alpha+\beta)},\\
    p_{\labelset_2} &= \sum_{(k,l)\in \labelset_2} \frac{1}{4}p_{kl}(\alpha,\beta,1) = 
    \frac{1}{2} - \frac{1}{2}e^{-4 \beta},\quad 
    p_D = \sum_{k=l} \frac{1}{4}p_{kl}(\alpha,\beta,1) = 
    \frac{1}{4} + \frac{1}{4}e^{-4\beta} + \frac{1}{2}e^{-2(\alpha+\beta)}.   
  \end{split}
  \]
  Therefore,
  \begin{equation}
    f_{\labelset_1} \asarrow p_{\labelset_1},
    f_{\labelset_2} \asarrow p_{\labelset_2}, \text{ and }
    f_D \asarrow p_D
    \label{k2p_limits}
  \end{equation}
  by the strong law of large numbers.
  We will need these a.s.\ limits when we express both \conv and \robust estimators in terms of 
  $f_{\labelset_1}$, $f_{\labelset_2}$, and $f_D$.
  \par
  The mle of $\gamma$,$\hat{\gamma} = -\frac{1}{4}\ln\left[1-\frac{4}{3}(1-f_D)\right]$,
  exists only if $1-f_D < 3/4$. Since we know that
  $1-f_D \asarrow \frac{3}{4} -\frac{1}{4} e^{-4\beta}-\frac{1}{2} e^{-2(\alpha+\beta)} < \frac{3}{4}$,
  we can safely assume that $\hat{\gamma}$ is well defined for large enough $n$.
  The \conv estimator 
  \begin{equation*}
    \text{E}_{\hat{\gamma}}(N_{\labelset_1}) = \hat{\gamma} \asarrow
    \frac{1}{4}\ln\left[1-\frac{4}{3}\left(\frac{3}{4}-\frac{1}{4}e^{-4\beta}-\frac{1}{2}e^{-2(\alpha+\beta)}
        \right)\right] = \beta - \frac{1}{4}\ln\left(\frac{1+2e^{2(\beta-\alpha)}}{3}\right).
  \end{equation*}
  To derive the limit of the \robust estimator we start with
  \begin{equation} 
    \frac{1}{n}\sum_{i=1}^n \text{E}_{\hat{\gamma}}\left(N_{\labelset_1}\,|\,\markovchain{0}=\obsdatascalar_{1i},
      \markovchain{1}=\obsdatascalar_{2i}\right)= 
    \sum_{k \in S} \sum_{l \in S} 
    \frac{f_{kl}}{p_{kl}(\hat{\gamma},1)}  
    \text{E}_{\hat{\gamma}}\left(N_{\labelset_1} 1_{\{\markovchain{1}=l\}}\,|\,\markovchain{0}=k\right).
    \label{jc_robust}
\end{equation}
Setting $\alpha=\beta$ in (\ref{k2p_prob}), we obtain transition probabilities for the Jukes-Cantor
model:
\begin{equation}
  p_{kl}(\gamma,t) = 
    \left(\frac{1}{4} - \frac{1}{4} e^{-4\gamma}\right) \indfun{k \ne l}+
    \left(\frac{1}{4} + \frac{3}{4} e^{-4\gamma}\right) \indfun{k = l}.
  \label{jc_prob}
\end{equation}
To get the functional form 
$\text{E}_{\hat{\gamma}}\left(N_{\labelset_1} 1_{\{\markovchain{1}=l\}}\,|\,\markovchain{0}=k\right)$,
we first notice that $\infgen^{\text{JC}}$ and $\infgen^{\text{JC}}_{\labelset_1}$ commute, leading to
\begin{equation*}
  \int_0^1 e^{\infgen^{\text{JC}}(\gamma)\tau} \infgen^{\text{JC}}_{\labelset_1}(\gamma)
  e^{\infgen^{\text{JC}(\gamma)}(1-\tau)}\text{d}\tau = 
  \infgen^{\text{JC}}_{\labelset_1}(\gamma)  e^{\infgen^{\text{JC}}(\gamma)} \int_0^1 \text{d}\tau = 
  \infgen^{\text{JC}}_{\labelset_1}(\gamma)  e^{\infgen^{\text{JC}}(\gamma)}. 
\end{equation*}
Hence, 
\begin{equation}
  \text{E}_{\hat{\gamma}}\left(N_{\labelset_1} 1_{\{\markovchain{1}=l\}}\,|\,\markovchain{0}=k\right)=
    \hat{\gamma}\left(\frac{1}{4} + \frac{3}{4} e^{-4\hat{\gamma}}\right) \indfun{(k,l) 
    \in \labelset_1} +
    \hat{\gamma}\left(\frac{1}{4} - \frac{1}{4} e^{-4\hat{\gamma}}\right) \indfun{(k,l) \notin \labelset_1}.
  \label{jc_counts}
\end{equation}
Plugging (\ref{jc_prob}) and (\ref{jc_counts}) to (\ref{jc_robust}), we arrive at
\begin{equation*}
  \begin{split}
    &\frac{1}{n}\sum_{i=1}^n \text{E}_{\hat{\gamma}}\left(N_{\labelset_1}\,|\,\markovchain{0}=\obsdatascalar_{1i},
      \markovchain{1}=\obsdatascalar_{2i}\right) = 
    \hat{\gamma}\left[1+4e^{-4\hat{\gamma}}\left(\frac{f_{\labelset_1}}{1-e^{-4\hat{\gamma}}}
        -\frac{f_D}{1+3e^{-4\hat{\gamma}}}\right)\right] \\
    &= -\frac{1}{4}\ln\left[1-\frac{4}{3}(1-f_D)\right]
    \times
    \left[1+\left(1-\frac{4}{3}(1-f_D)\right)\left(\frac{3 f_{\labelset_1}}{1-f_D}-1\right)\right].
  \end{split}
\end{equation*}
Plugging in limits (\ref{k2p_limits}) in the above formula produces the desired result.
\end{proof}

\begin{proof}[Proof of Corollary \ref{limit_cor}]
  Defining   
  $A = (e^{-4\beta} + 2e^{-2(\alpha+\beta)})/3$, we write the limiting difference of the \robust and 
  \conv estimates as
  \begin{equation}
    \label{est_diff}
    \mu^{im}_{\infty} - \mu^{pi}_{\infty} = -\frac{A \ln A}{3(1-A)}e^{-4\beta}(1-e^{-2(\alpha-\beta)}).
  \end{equation}
  Since $0<A \le 1$, $\mu^{im}_{\infty} - \mu^{pi}_{\infty}$ and $\alpha - \beta$ always have the same sign.
  Moreover, using $e^x \ge 1+x$, we can show that 
  \begin{equation}
    \label{A_ineq}
    0 < - \frac{A \ln A}{1-A} = \frac{\ln(1/A)}{1/A -1} < 1,
  \end{equation}
  when $\alpha \ne \beta$.
  Recall that $\mu = \statdist^T \infgen^{K2P}(\alpha,\beta) \mathbf{1} = \alpha$.  leading to
  \begin{equation*}
    \mu^{pi}_{\infty} - \mu = \beta - \frac{1}{4}\ln\left(\frac{1+2e^{2(\beta-\alpha)}}{3}\right) - \alpha = 
    -\frac{1}{4}\ln\left(\frac{e^{4(\alpha-\beta)}+2e^{2(\alpha-\beta)}}{3}\right).
  \end{equation*}
  Hence, $\mu^{pi}_{\infty} - \mu$ and $\alpha - \beta$ always have opposite signs. 
  \paragraph{Case 1: $\alpha > \beta$.} We have $0 < e^{-4\beta}(1-e^{-2(\alpha-\beta)}) < 2(\alpha - \beta)$, 
  which together with \ref{A_ineq} imply
  $0 < \mu^{im}_{\infty} - \mu^{pi}_{\infty} < \frac{2}{3}(\alpha - \beta)$.
  Next, we use concavity of logarithm and arrive at 
  $\mu^{pi}_{\infty} - \mu < (\alpha-\beta)/3$. 
  Combining these last two inequalities, we have $\mu^{im}_{\infty} - \mu < 0$. Therefore, 
  \begin{equation*}
    \mu^{pi}_{\infty} - \mu = \mu^{pi}_{\infty} - \mu^{im}_{\infty} + \mu^{im}_{\infty} - \mu < \mu^{im}_{\infty} - \mu < 0,
  \end{equation*}
  which proves the desired inequality.
  \paragraph{Case 2: $\alpha < \beta$.} Recall that $\mu^{pi}_{\infty} > \mu$. Plugging in 
  $0 > e^{-4\beta}\left(1-e^{-2(\alpha-\beta)}\right) = 
  e^{-2(\alpha+\beta)}\left(e^{2(\alpha - \beta)}-1\right) > e^{2(\alpha - \beta)}-1$
  to (\ref{est_diff}) we arrive at 
    $\frac{1}{3}\left(e^{2(\alpha - \beta)}-1\right) < \mu^{im}_{\infty} - \mu^p_{\infty} < 0$. So
  \begin{equation*}
    \mu^{im}_{\infty} - \mu = \mu^{im}_{\infty} -\mu^{pi}_{\infty} + \mu^{pi}_{\infty} - \mu >
    \frac{1}{3}\left(e^{2(\alpha - \beta)}-1\right) 
    -\frac{1}{4}\ln\left(\frac{e^{4(\alpha-\beta)}+2e^{2(\alpha-\beta)}}{3}\right).
  \end{equation*}
  Defining the function on the right-hand side of the above inequality as 
  $w(\alpha-\beta)$, we show that $w(0) = 0$ and 
  \[
  w'(\delta) = \frac{2}{3}e^{2\delta} - \frac{3\left(e^{2\delta}+1\right)}
  {e^{2\delta} + 2} = \frac{\left(e^{2\delta}-1\right)\left(2e^{2\delta}+3\right)}
  {3\left(e^{2\delta}+2\right)} < 0
  \]
  for $\delta < 0$. Therefore, we have $\mu^{im}_{\infty} - \mu > w(\alpha - \beta) > 0$ and 
  \[
  \mu^{pi}_{\infty} - \mu = \mu^{pi}_{\infty} - \mu^{im}_{\infty} + \mu^{im}_{\infty} - \mu > 
  \mu^{im}_{\infty} - \mu > 0.  
  \]

\end{proof}








\end{document}